\documentclass[prb,twocolumn]{revtex4-2}

\usepackage{verbatim}
\usepackage{bm}
\usepackage{amstext}
\usepackage{amssymb}
\usepackage{stackrel}
\usepackage{graphicx}

\begin{document}
\title{Charge trapping and recombination in dipolar field of charged defect
cluster in silicon}
\author{Darius Abramavicius}
\affiliation{Institute of Chemical Physics, Physics Department, Vilnius University, Sauletekio al. 9-III}
\author{Juozas Vidmantis Vaitkus}
\affiliation{Institute of Photonics and Nanotechnology, Physics Department, Vilnius University, Sauletekio
al. 9-III}

\begin{abstract}
Extensive irradiation of silicon crystal sensors by high energy particles
in e. g. accelerators yield defect clusters of different types. Trapping
of electrons and holes result in extended internal electric fields
that drive remaining free charges. The question whether these internal
electric fields affect the experimental observables, e.g. recombination process and lifetime of free
charges is the main focus of this paper. Including the drift and diffusion
of electrons and holes we calculate the recombination rate in a cubic
sample with a single dipolar cluster of defects. It is shown that
 the large effect on charge lifetime is to be expected when charge
diffusion length during the charge lifetime is comparable to
the dimensions of the cluster. If the diffusion length exceeds the cluster size,  the cluster barely affects
the recombination rate. 
\end{abstract}
\maketitle

\section{Introduction}

Silicon-based semiconductor sensors and detectors are widely used as high energy
(HE) particle detectors \citep{Hartmannbook}. The fundamental principal
design of detectors relies on generation of mobile charges by HE particles.
These mobile charges are being extracted and detected as electrical
signals in e. g. avalanche diode configurations, such as low gain
avalanche detectors (LGADs)\citep{Moffat2018}. However, in specific
experiments the irradiation energy may soar as high as 13-14 TeV (in
e. g. Large Hadron Colliders - LHCs at CERN); the HE particles strongly
interact with the crystal lattice and their kinetic energy is sufficient
to knock-out atoms from their rest positions. In the simplest case
two types of structural defects are then generated in the bulk: vacancies
and interstitials, which shorten the devices' lifetime and the detectors
degrade \citep{Lindstrom2003,Gallrapp2017}. It has been proposed
that atoms, knocked out from their lattice positions have sufficient
energy to further on knock out other atoms, thus creating an avalanch-like
defect generation of vacancies and interstitial atoms \citep{Radu2015}.
Additional to that, as interstitials are created in the lattice they
could relax into several distinct lattice positions. The interstitial
defects tend to migrate in the crystal, recombine with vacancies until
the defects relax into quasi-stable configuration after impact with
the HE particle. As a result, the complex pattern of extended defect clusters
should be highly probable \citep{Radu2015,Centoni2005}. Clustering of defects
is also induced by ablation process \citep{Aye2021}. It is highly
probable that groups of interstitials and vacancies organize into
nano-size effective clusters where high content of vacancies and interstitials
are distributed non homogeneously \citep{Radu2015}. Resulting defect
clusters may demonstrate dipolar-like characteristics.

At least three types of interstitials are stable in Si crystal \citep{Newman1982,Leung1999,Centoni2005}
with their unique electronic properties. Electronic properties of
point defects are essentially determined by position of electron energy
at the defect \citep{McCluskeybook,McCluskey2020}. Some of the point defects can trap electrons,  others
can trap holes, while some are recombination centers \citep{Cottom2016}. Theoretical determination
of isolated point defects can be calculated by quantum methods at high accuracy 
at the atomic level \citep{Alkauskas2016}. Clustering of various defects becomes
impossible to analyze at the same level.
Nonhomogeneous distribution of point defects in defect cluster accompanied
by trapping of charges creates long-range internal lattice fields
that affects dynamics of the remaining charges on long distances \citep{Zasinas2016}.
Strong correlation between the type of defect clustering and the type
of irradiation has been established \citep{Huhtinen2002,Aye2021}.
Defect clusters are most probable after irradiation by neutrons, while
irradiation by charged particles may lead to sparse point
defects \citep{Radu2015}.  Wide distribution of types of defect
clusters should be thus expected. It has been estimated that only
about a third of point defects become polarized, i. e. trap one or
other type of charges, while the point defects that effectively trap
both charges, become recombination centers \citep{Vaitkus2021}. Hence,
the largest amount of defects turn into recombination centers, which
can be assumed to be distributed uniformly in the vicinity of (and
within) the cluster. In the simplified picture, the charged defects
thus create  polarized medium for free
charges and it thus becomes a significant factor, which drives
the free charges that become captured by neutral recombination centers.
The question, whether such internal fields affect the charge recombination
and at what conditions, is targeted in this paper. 

Experimental observations of recombination kinetics after photoexcitation
allow estimation of charge recombination rates and demonstrate that
the charge recombination time becomes directly proportional to the
overall doze of irradiation \citep{Gaubas2008,Zasinas2016}. It is
still debatable whether this variation of charge lifetime is due to
increase of overall defect density in the bulk or due to spatial charge
accumulated dipolar-like defect clusters because presumably 
they are responsible for variation of charge lifetime as the function of 
irradiation intensity \citep{Gaubas2018}. In this work we focus on
this question by computer simulations of the recombination process
by considering diffusion and drift processes of free charges in the
vicinity of the simplest possible defect cluster with the restriction
that the density of point defects is assumed to be not too high so
that the bulk properties, such as mobility and diffusion coefficients
of charges are not affected. This is certainly an approximation, however, it allows to highlight the effect
that 
the dipolar field of the defect cluster makes on the charge recombination.   Simulations of the recombination process
in a box with a single  cluster demonstrate that due to large
diffusion distances of free electrons and holes, a nano-sized defect
clusters with dipolar internal fields weakly affect charge recombination
rates when the density of recombination centers is low. Influence
of the  internal fields becomes considerable when the charge lifetime is artificially reduced
 so that the diffusion length 
becomes comparable to the size of the cluster. 

\section{Theory of charge density evolution }

Consider a single defect cluster that is isolated in the crystal.
Electron-hole recombination process in the defect cluster involves
(at least) two opposite free charges, both mobile, both affected by the internal
field of trapped charges. Their dynamics  has to be
followed simultaneously. They must interact via Coulomb field and
with the field of trapped charges in the defect cluster. We restrict
the problem to the semiclassical approach in the overdamped limit
(relaxation-time approximation), where the charges are characterized
by their effective masses and relaxation times. The main parameters
characterizing their dynamics are the diffusion coefficients and mobilities. 

A pair of free charges can be described in space by a distribution
function, $\rho(\bm{r}_{e},\bm{r}_{h},t)d^{6}r$ in six-dimensional
space, denoting the probability that electron is at position $\bm{r}_{e}$
in the volume element $d^{3}r$, while the hole is at $\bm{r}_{h}$
in volume element $d^{3}r$. Dynamics is governed by diffusion and drift processes
described by Fokker-Planck equation of motion \citep{Breuerbook2007}:
\begin{eqnarray}
\dot{\rho}(\bm{r}_{e},\bm{r}_{h},t) & = & D_{e}\Delta_{e}\rho(\bm{r}_{e},\bm{r}_{h},t)+D_{h}\Delta_{h}\rho(\bm{r}_{e},\bm{r}_{h},t)\nonumber \\
 &  & -\nabla_{e}\cdot\left(\rho(\bm{r}_{e},\bm{r}_{h},t)\bm{f}_{e}(\bm{r}_{e},\bm{r}_{h},t)\right)\nonumber \\
 &  & -\nabla_{h}\cdot\left(\rho(\bm{r}_{e},\bm{r}_{h},t)\bm{f}_{h}(\bm{r}_{e},\bm{r}_{h},t)\right)\nonumber \\
 &  & -S(\bm{r}_{e},\bm{r}_{h})\rho(\bm{r}_{e},\bm{r}_{h},t).\label{eq:Full6D}
\end{eqnarray}
 Here the dot denotes time derivative, we also assume the homogeneous diffusion with diffusion coefficients
$D_{e}$ for electrons and $D_{h}$ for holes,  $\bm{f}_{e}(\bm{r}_{e},\bm{r}_{h},t)$
is the drift force acting on electrons, while $\bm{f}_{h}(\bm{r}_{e},\bm{r}_{h},t)$
is the corresponding force acting on holes. Correspondingly, $\Delta_{i}$
and $\nabla_{i}$ are the 3D Laplace and \emph{nabla} differential
operators acting on electron and hole coordinates, respectively. $S(\bm{r}_{e},\bm{r}_{h})$
describes the recombination. All these parameters are described below. 

The combined probability density can be used to define separate electron
and hole distributions. We can define the electron distribution function
by integrating out the hole variables
\begin{equation}
\rho_{e}(\bm{r}_{e},t)=\int_{V}d^{3}\bm{r}_{h}\rho(\bm{r}_{e},\bm{r}_{h},t);
\end{equation}
and similarly for the hole distribution function 
\begin{equation}
\rho_{h}(\bm{r}_{h},t)=\int_{V}d^{3}\bm{r}_{e}\rho(\bm{r}_{e},\bm{r}_{h},t),
\end{equation}
 while the total probability (or density) of electrons and holes being in the system
is given by
\begin{equation}
n(t)=\int_{V}d^{3}\bm{r}_{h}\int_{V}d^{3}\bm{r}_{e}\rho(\bm{r}_{e},\bm{r}_{h},t).
\end{equation}
 The total probability density $n(t)$ is the quantity of interest
when characterizing the charge recombination process. Following time
evolution of $n(t)$ allows to define the recombination rate. 

Above definitions yield separate, while coupled, equations for electron
and hole distributions by integrating equation \ref{eq:Full6D}. Using
notation $(i,-i)=(e,h)$ or $(h,e)$ we find 
\begin{eqnarray}
\dot{\rho}_{i}(\bm{r},t) & = & D_{i}\Delta\rho_{i}(\bm{r},t)\nonumber \\
 &  & -\nabla_{r}\cdot\left(\int_{V}d^{3}\bm{r}'\rho_{i}(\bm{r},t)\frac{\rho_{-i}(\bm{r}',t)}{n(t)}\bm{f}_{i}(\bm{r},\bm{r}',t)\right)\nonumber \\
 &  & -\int_{V}d^{3}\bm{r}'S(\bm{r},\bm{r}')\frac{\rho_{-i}(\bm{r}',t)}{n(t)}\rho_{i}(\bm{r},t).\label{eq:FP-final}
\end{eqnarray}
 Here the first term on the right side of equation is due to diffusion,
second is responsible for electron-hole interaction and interaction
with other trapped charges, while the third term defines the recombination
process. 

To define explicit expression for the drift forces we recall that the drift
force in Fokker-Planck equation corresponds to equations of motion
for a particle with coordinate $x_{i}(t)$ in the form
\begin{equation}
\dot{x}_{i}(t)=f_{i}(x_{1}...x_{N},t).\label{eq:standard-eqm-form}
\end{equation}
The force function $f_{i}(x_{1}...x_{N},t)$ can be defined by considering
all internal Coulomb fields. For electrons and holes in the semiclassical
approximation we have to include inter-particle Coulomb attraction
as well as internal field-induced attraction or repulsion by the trapped charges. All these fields
may be expressed as gradients of internal potential created by charged
species:
\begin{eqnarray} \label{eq:coulomb}
U(\bm{r}_{e},\bm{r}_{h}) & = & -\chi\frac{1}{\left|\bm{r}_{e}-\bm{r}_{h}\right|}\nonumber \\
 &  & -\chi\sum_{d}Q_{d}\left(\frac{1}{\left|\bm{r}_{e}-\bm{R}_{d}\right|}-\frac{1}{\left|\bm{r}_{h}-\bm{R}_{d}\right|}\right)
\end{eqnarray}
 Here the first term corresponds to the electron-hole attraction of mobile charges
with $\chi=e^{2}/(4\pi\epsilon\epsilon_{0})$ being the amplitude
of the Coulomb potential, other two terms are due to static trapped charges: $Q_{i}$ is the charge of the defect $d$
given in terms of elementary electron charge $e$, while $\bm{R}_{d}$
- its position. Considering the times longer than the scattering time,
averaged motion of the charges in the overdamped limit can be described
by equation in the form of eq \ref{eq:standard-eqm-form} with $i=e,h$,
what defines the drift for $i-$ th particle \citep{ashcroftbook}
\begin{equation}
\dot{\bm{r}}_{i}=-\frac{\tau_{i}}{m_{i}^{\ast}}\nabla_{i}U(\bm{r}_{i},\bm{r}_{-i})\equiv\bm{f}_{i}(\bm{r}_{e},\bm{r}_{h}).\label{eq:drift-force}
\end{equation}
Here we find $m_{i}^{\ast}$ - the effective mass, $\tau_{i}$ is
the relaxation  time. Also notice that the parameters, which enter
the drift force in Eq. \ref{eq:drift-force} can be related via Drude
expression, relating the relaxation time, effective mass, and mobility:
$\mu_{i}=e\tau_{i}/m_{i}^{\ast}$; $e$ is the electron charge. Consequently,
the ratio $\tau_{i}/m_{i}^{\ast}$ reduces to $\tau_{i}/m_{i}^{\ast}=\mu_{i}/e$.
Then Einstein relation $D_{i}=\mu_{i}k_{B}T$ allows to reduce the
drift amplitude to the thermal energy. Further on we use the standard
notation for the inverse thermal energy $\beta=(k_{B}T)^{-1}$ and
can write $\tau_{i}/m_{i}^{\ast}=\beta D_{i}/e$.

The last term in eq. \ref{eq:FP-final} characterizes the recombination.
Assuming that \emph{the process} is time independent and local in space,
i. e. the electron and hole must meet in space on a specific recombination
center in order the recombination to occur; and that the probability
of finding a recombination center is distributed uniformly in space,
we obtain a simple expression for the recombination function $S(\bm{r}_{e},\bm{r}_{h})\equiv\bar{R}\delta(\bm{r}_{e}-\bm{r}_{h})$,
with $\bar{R}$ - the mean on-site recombination rate.  

Combining the recombination rate and drift force into Eq. \ref{eq:FP-final}
we obtain the  real-space equations of motion describing drift, diffusion
and recombination of electrons and holes in the cluster:
\begin{eqnarray}
 &  & \dot{\rho}_{i}(\bm{r},t)=D_{i}\Delta\rho_{i}(\bm{r},t)-\bar{R}\rho_{i}(\bm{r},t)\frac{\rho_{-i}(\bm{r},t)}{n(t)}\nonumber \\
 &  & \qquad-\frac{\beta D_{i}\chi}{e}\int_{V}d^{3}\bm{r}'\frac{\rho_{-i}(\bm{r}',t)}{n(t)}\frac{\bm{r}-\bm{r}'}{\left|\bm{r}-\bm{r}'\right|^{3}}\cdot\nabla\rho_{e}(\bm{r},t)\nonumber \\
 &  & \qquad+\frac{\beta D_{i}\chi}{e}q_{i}Q_{d}\sum_{d}\frac{\bm{r}-\bm{R}_{d}}{\left|\bm{r}-\bm{R}_{d}\right|^{3}}\cdot\nabla\rho_{e}(\bm{r},t)\label{eq:real-space-final}
\end{eqnarray}
where $q_{i}$ is the charge of the $i-$th particle in terms of elementary
electron charge. Integration volume $V$ corresponds to the system
under consideration. Notice that resulting equations are linear in
electron and hole density so the absolute density does not play any
role. 

In the following we study the recombination process in the simplest
neutral system, where the internal field is of the dipolar type that
can be represented by positioning two opposite stationary charges
at the given positions. It is convenient to set orientation of the
coordinate system with respect to the dipole: the
symmetry axis of the dipole is oriented along $z$ axis, positive charge of
the dipole is positioned at $\bm{z}_{0}\equiv(0,0,z_{0})$, while
negative - at $-\bm{z}_{0}$. The absolute charge center is then at the origin. 

\section{Stationary equation in reciprocal space}

The obtained equations of motion depend on spatial as well as on time
derivatives. Numerically the high order mixed derivatives are poorly
represented by a finite difference. Certain derivatives can be eliminated
by transforming the problem into the reciprocal $\bm{k-}$ space by
applying Fourier transform. Using definition 
\begin{equation}
\rho_{i}(\bm{r},t)=\int\frac{d^{3}\bm{k}}{(2\pi)^{3}}e^{i\bm{k}r}\sigma_{i}(\bm{k},t)
\end{equation}
 we have a simple conversion for the core of the Coulomb potential:
\begin{equation}
\frac{1}{|\bm{r}|}\to\frac{4\pi}{|\bm{k}|^{2}}.
\end{equation}
   Additionally, at zero wavevector $\sigma_{i}(0,t)=\sigma_{-i}(0,t)=n(t)$
is the total charge density. Consequently, evaluating the recombination
kinetics becomes equivalent to following the zero-value wave vector
amplitude. The resulting equation in $\bm{k}-$ space becomes simplified:
\begin{eqnarray}
 &  & \dot{\sigma}_{i}(\bm{k},t)=-\bm{k}^{2}D_{i}\sigma_{i}(\bm{k},t)\nonumber \\
 &  & \ +\frac{CD_{i}L}{8\pi^{3}n(t)}\int d^{3}\bm{k}'\frac{\bm{k}\cdot\bm{k}'}{|\bm{k}'|^{2}+\eta^2}\sigma_{-i}(\bm{k}',t)\sigma_{i}(\bm{k}-\bm{k}',t)\nonumber \\
 &  & \ +\frac{iq_{i}QCD_{i}L}{4\pi^{3}}\int d^{3}\bm{k}'\frac{\bm{k}\cdot\bm{k}'}{|\bm{k}'|^{2}+\eta^2}\sin\left(\bm{k}'\bm{z}_{0}\right)\sigma_{i}(\bm{k}-\bm{k}',t)\nonumber \\
 &  & \ -\frac{\bar{R}}{8\pi^{3}n(t)}\int d^{3}\bm{k}'\sigma_{-i}(\bm{k}',t)\sigma_{i}(\bm{k}-\bm{k}',t).\label{eq:k-final}
\end{eqnarray}
 Here we denoted $C=4\pi\beta\chi/(eL)$ - the dimensionless temperture-dependant
amplitude of the Coulomb interaction. The first term in the right
hand side of Eq. \ref{eq:k-final} is the diffusion-related damping,
then the integrals are due to Coulomb drift force and final is the
recombination term. The drift-related integral vanishes
for $\bm{k}=0$ or $\bm{k}'=0$ since the corresponding amplitudes
represent homogeneous charge densities having translational invariance.
To emphasize this and to avoid divergencies, an additional $\eta^{2}$
shift with $\eta\ll\Delta k$ is added in the denominators
 (the Coulomb potential function in real
space becomes equal to the Yukawa potential).

While equations do not involve spatial derivatives (i. e. diffusion
has been transformed into derivative-less form), the tripple integrations
over momentum $\bm{k}'$ cannot be reduced, so full numerical propagation
at every time is computationally very expensive. Additionally, notice
that the diffusion-related terms in the equation are proportional
to $\bm{k}^{2}$. So the diffusion-related term diverges for large
wave vectors requiring decreasing of time step. 
To avoid explicit time propagation we consider the stationary dynamics,
when the charge distribution
relaxed in space for the given steady
decay of nonstationary populations. Then the recombination rate can
be evaluated with the Ansatz that the electron (and hole) density
decays exponentially. We thus request 
\begin{equation}
\sigma_{i}(\bm{k},t)=\bar{\sigma}_{i}(\bm{k})e^{-\gamma t}
\end{equation}
 With the initial condition that at time zero, the probability to
find electron and hole is equal to $1$, we thus have
the initial condition corresponding to the boundary value  $\bar{\sigma}_{i}(\bm{k}=0)\equiv n(t=0)=1$.
For $\bm{k}\neq0$ from Eq. \ref{eq:k-final} we obtain the integral
equation for the spatial charge distribution in discretized form
\begin{eqnarray}
 &  & \bar{\sigma}_{i}(\bm{k})=\frac{C}{(\bm{k}L)^{2}}\sum_{\bm{k}'}\frac{\bm{k}\cdot\bm{k}'}{|\bm{k}'|^{2}}\bar{\sigma}_{-i}(\bm{k}')\bar{\sigma}_{i}(\bm{k}-\bm{k}')\label{eq:final}\\
 &  & \qquad+\frac{2iq_{i}QC}{(\bm{k}L)^{2}}\sum_{\bm{k}'}\frac{\bm{k}\cdot\bm{k}'}{|\bm{k}'|^{2}}\sin\left(\bm{k}'\bm{z}_{0}\right)\bar{\sigma}_{i}(\bm{k}-\bm{k}')\nonumber \\
 &  & \qquad+\frac{\gamma}{\bm{k}^{2}D_{i}}\bar{\sigma}_{i}(\bm{k})-\frac{\gamma_{h}}{\bm{k}^{2}D_{i}}\sum_{\bm{k}'}\bar{\sigma}_{-i}(\bm{k}')\bar{\sigma}_{i}(\bm{k}-\bm{k}'),\nonumber 
\end{eqnarray}
 while $\bm{k}=0$ in Eq. \ref{eq:k-final} yields the decay rate
of the charge density
\begin{eqnarray}
\gamma & = & \gamma_{h}\sum_{\bm{k}'}\bar{\sigma}_{-i}(\bm{k}')\bar{\sigma}_{i}(-\bm{k}').\label{eq:rate-final}
\end{eqnarray}
 We also introduced $\gamma_{h}=\bar{R}/V$, being the recombination rate for the homogeneous system (see next section). In the discretized form all terms 
 are dimensionless. These equations can be solved iteratively starting
from e. g. the homogeneous distribution.  

\section{Results}

Homogeneous distribution of charges constitutes the reference point
which represents the case when the internal defect field is not present.
In that case the combined 6D normalized density can be written as
\begin{equation}
\rho^{(hom)}(\bm{r}_{e},\bm{r}_{h},t)=\frac{n^{(hom)}(t)}{V^{2}},
\end{equation}
where $V$ is the volume under consideration. Consequently 
\begin{equation}
\rho_{e}^{(hom)}(\bm{r},t)=\rho_{h}^{(hom)}(\bm{r},t)=\frac{n^{(hom)}(t)}{V}  
\end{equation}
 and in reciprocal space 
 $\sigma_{i}^{(hom)}(\bm{k},t)=n^{(hom)}(t)$
for $\bm{k}=0$ and zero otherwise. For the homogeneous system the
translational invariance holds (all Coulomb interactions vanish).
Equation \ref{eq:k-final} can then be integrated what yields the
recombination rate, $\gamma_{h}=\bar{R}/V$, of the homogeneous distribution
\begin{equation}
\frac{\dot{n}^{(hom)}(t)}{n^{(hom)}(t)}=-\gamma_{h}.\label{eq:homogeneous_rate}
\end{equation}
 Notice that it rescales the on-site rate $\bar{R}$ by the volume
of the system. This is related to the fact that when the volume of
the system is increased, the probability for the two carriers to meet
on the single site, which is proportional to $V^{-2}$, is reduced.
The scaling in Eq. \ref{eq:homogeneous_rate} is $V^{-1}$ because
for the constant density of recombination centers, their number increases
linearly with volume $V$. However, in our simulations the rate $\gamma_{h}$
should be considered as tuning parameter since the density of recombination
centers in defect clusters is not known.

The homogeneous distribution function is taken as the initial
condition for numerical solution of Eq. \ref{eq:final}. Starting
from $\bar{\sigma}_{i}^{(hom)}(\bm{k}=0)=1$ (and zero otherwise) we solved Eq. \ref{eq:final}
iteratively together with Eq. \ref{eq:rate-final}. For calculations
we assume the widely available set of parameters of pure Si semiconductor. 
We used a length scale of 1$\mu m$ and time scale of $1\mu s$.
The modelling system is taken as a box of volume $V=L^{3}=1\,(\mu m)^{3}$.
 The parameters, that enter the drift force in Eq. \ref{eq:drift-force},
specifically, the effective masses and relaxation times, have been
reduced to  diffusion constants and mobilities and finally to thermal
energy in Eq. \ref{eq:final}. We use the standard set of Si
characteristics at room temperature, i. e. the experimental mobility
values, $\mu_{e}=1400\ cm^{2}/(Vs)$, $\mu_{h}=450\,cm^{2}/(Vs)$,
and diffusion constants, $D_{e}=36\,cm^{2}/s$, $D_{h}=12\,cm^{2}/s$.
Additionally the Coulomb field is scaled by Si dielectric constant
$\epsilon=11.68$.  The box is discretized
into $40\times40\times40$ points. The resulting effective cubic lattice
constant is $a=1/40$ $\mu m$. In the reciprocal space the 
integration step $\Delta k=2\pi/L=2\pi$ $(\mu m)^{-1}$.

The last parameter that has to be considered is the trapped charge
in the defect $Q$. For the defect cluster we assume that the cluster
contains net charge $Q>e$, i. e. the defect cluster can trap more
then a single charge. The total trapped charge is the property of
the defect and is equal to the number of trapping sited in the cluster.
A single point defect can in principle accommodate only 1-2 e. 
However, in the extended defect cluster we may consider accumulation 
of several point defects in a single region. 
Consequently,  the charge $Q$ becomes a unique number for each particular cluster (Eq. \ref{eq:final}).
In present study we chose this number arbitrarily and set $Q=4e$.

\begin{figure}
\begin{centering}
\includegraphics[width=1\columnwidth]{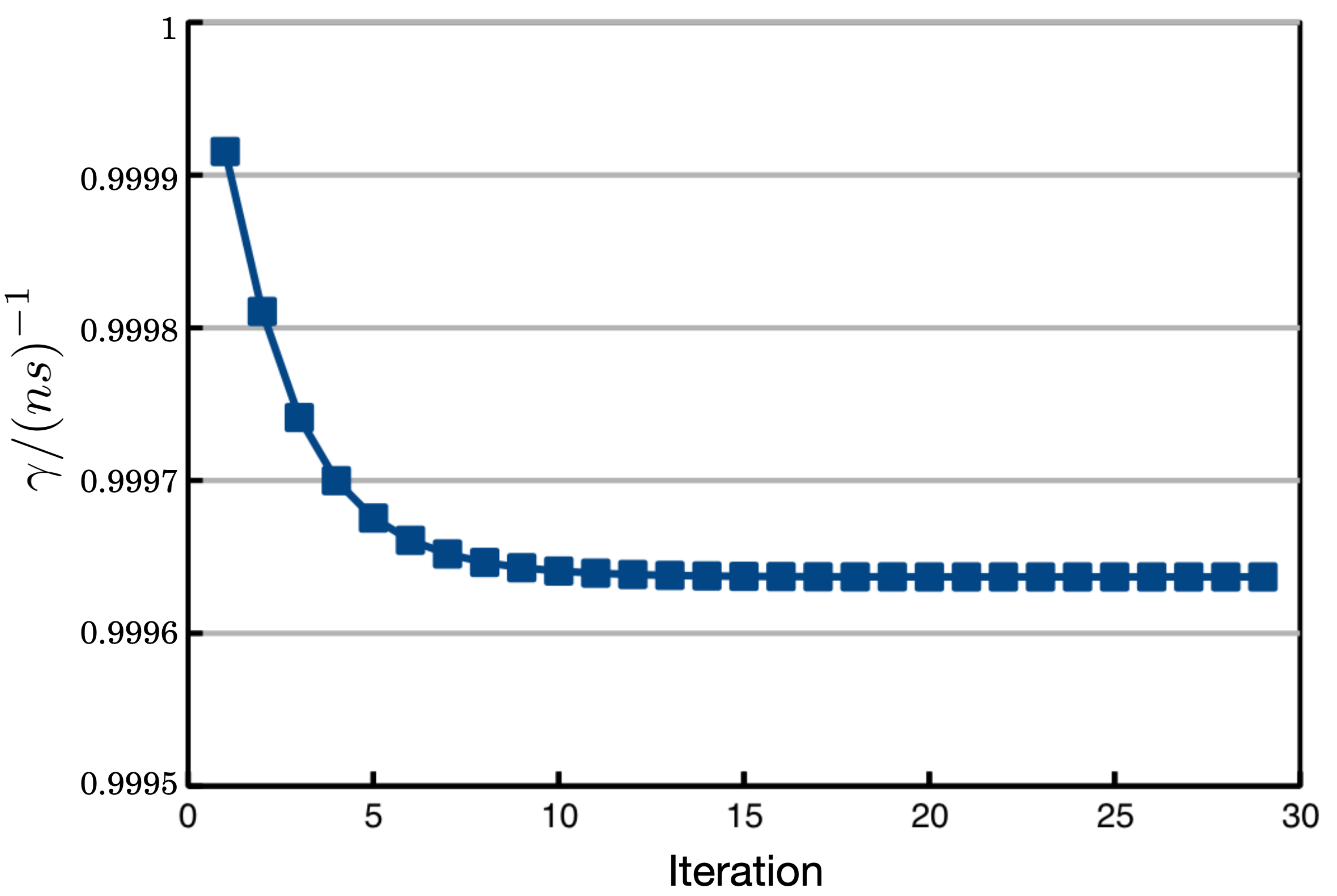}
\par\end{centering}
\caption{\label{fig:iteration}Convergence of decay rate with
iterations. Dipole charges $Q=4e$, dipole separation $2z_{0}=4a\equiv100$
nm, homogeneous recombination rate $\gamma_{h}=1$ ns$^{-1}$.}
\end{figure}

\begin{figure}
\begin{centering}
\includegraphics[width=1\columnwidth]{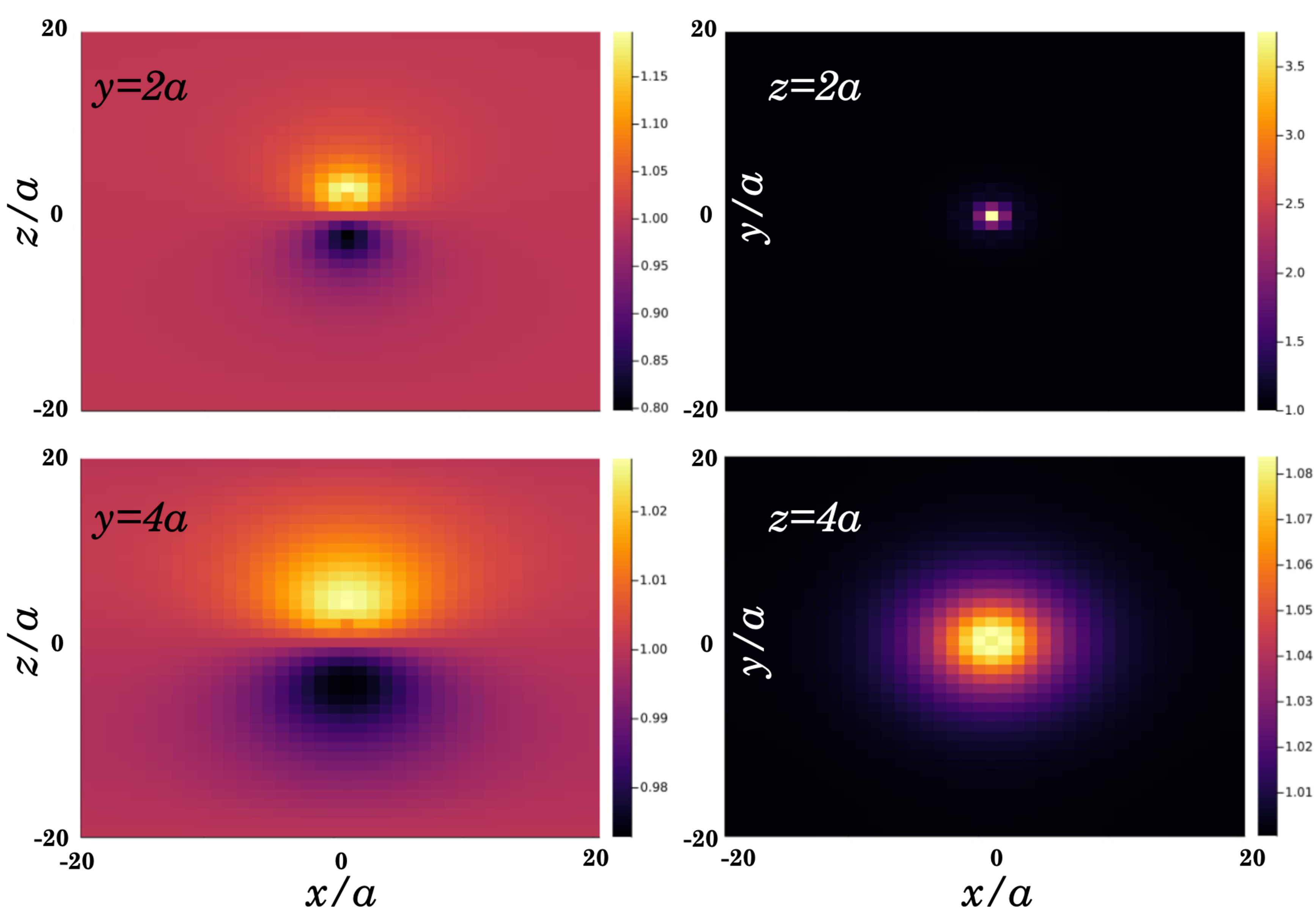}
\par\end{centering}
\caption{\label{fig:Distribution-of-electron-in-plain}Distribution of electron
density in real space $\bar{\sigma}_{e}(\bm{r})$ at a specific plain. Dipole
charges $Q=4e$, dipole separation $2z_{0}=4a\equiv100$ nm, homogeneous
recombination rate $\gamma_{h}=1$ ns$^{-1}$.}

\end{figure}

\begin{figure}
\begin{centering}
\includegraphics[width=1\columnwidth]{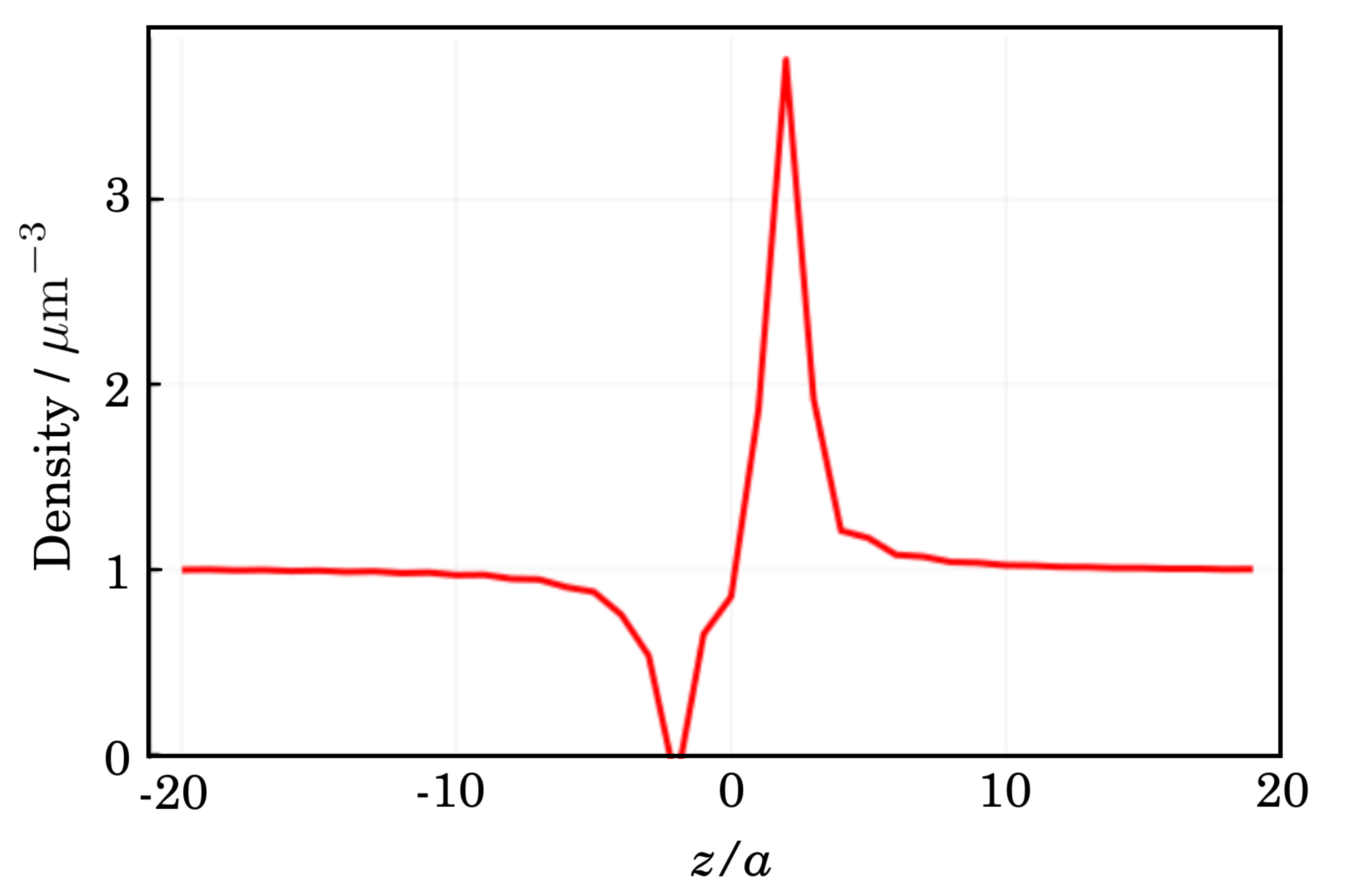}
\par\end{centering}
\caption{\label{fig:Distribution-of-electron}Distribution of electron density
$\bar{\sigma}_e(\bm{r})$ along $z$ axis at $x,y=0$. Dipole charges
$Q=4e$, dipole separation $2z_{0}=4a\equiv100$ nm, homogeneous recombination
rate $\gamma_{h}=1$ ns$^{-1}$. }
\end{figure}

\begin{figure}
\begin{centering}
\includegraphics[width=1\columnwidth]{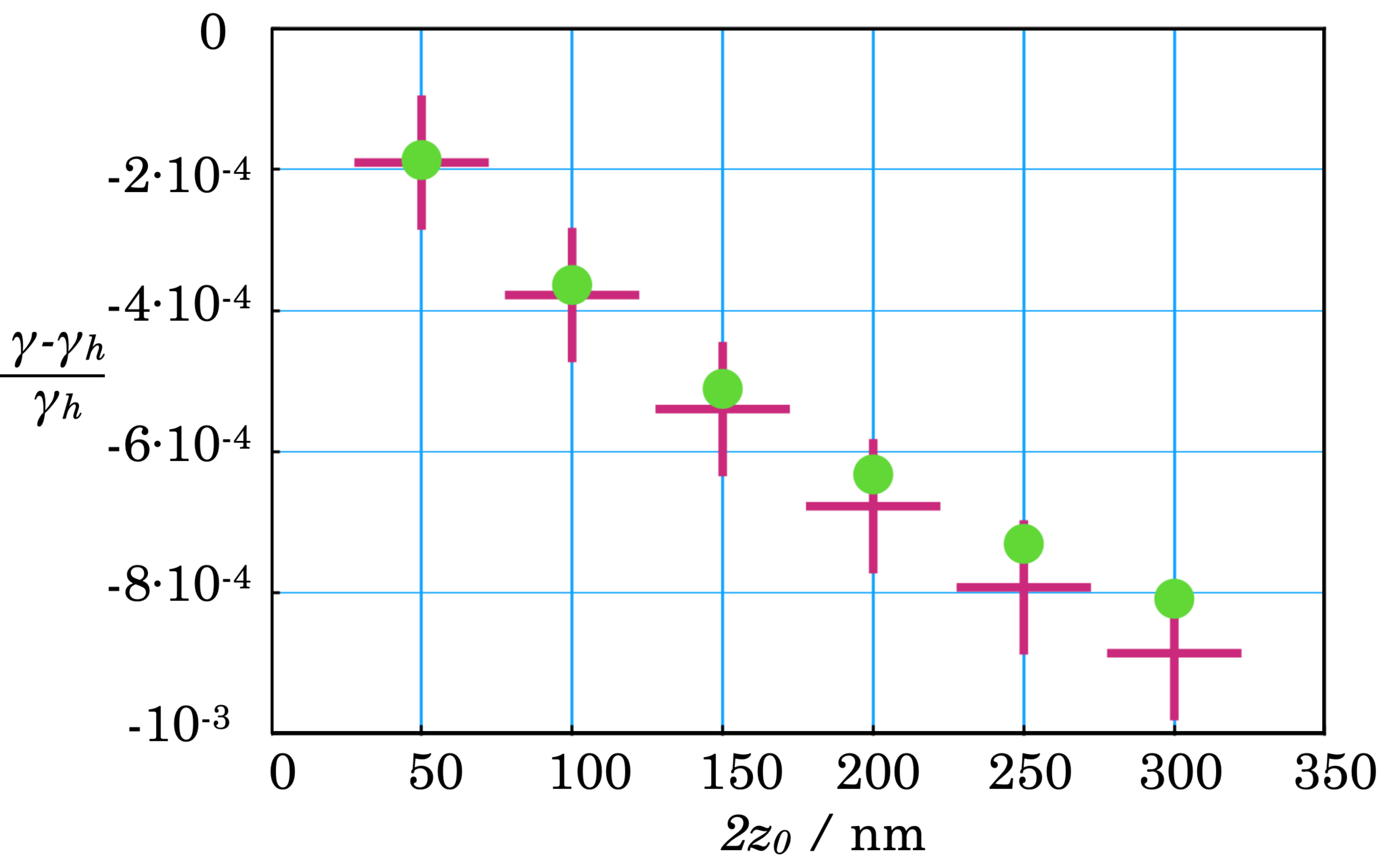}
\par\end{centering}
\caption{\label{fig:relaxation_rates}Normalized variation of recombination
rates for dipole charges $Q=4e$ as a function of separation between
dipole charges when $\gamma_{h}=1$ ns$^{-1}$ (green dots) and (100
ps)$^{-1}$ (red crosses). }
\end{figure}

First, we consider the  slow recombination regime when the
probability of recombination centers is low. This is the case when
for our chosen size of the cluster we take $\gamma_{h}<(\Delta k)^{2}D_{i}$
in Eq. \ref{eq:final}. This corresponds to the case when the diffusion-related
charge travel time over the system length $L$ is shorter than the
recombination time $\gamma_{h}^{-1}$. For the electron and hole the diffusion coefficients
we find $(\Delta k)^{2}D_{e}=$ $142$ ns$^{-1}$ for electrons and
$(\Delta k)^{2}D_{h}=$ $47$ ns$^{-1}$ for holes. To represent this
case we assume that the homogeneous recombination rate parameter $\gamma_{h}=1$
ns$^{-1}$. Fig. \ref{fig:iteration} demonstrates iterative convergence
of Eq. \ref{eq:final} towards the solution for this type of system.
The convergence is mostly exponential and can be safely truncated after
$\sim$20-th iteration. The resulting electron distribution is presented
in Fig. \ref{fig:Distribution-of-electron-in-plain} at the plains
$xy$, $xz$, while along $z$ axis at $(x,y)=0$ is shown in Fig
\ref{fig:Distribution-of-electron}. As expected, the positive trapped
charge attracts the electron density, while the negative trapped charge
repels it. The distribution becomes rotationally symmetric around
$z$ axis. It is worth noting that due to relatively large diffusion
constants, the distribution is much wider at distances larger than
the dipolar charge separation. The integral charge around the dipole
(while with low amplitude) makes significant contribution to the recombination
process, but the conditions are essentially quite similar to the homogeneous
conditions and the effect of the dipolar field of the trapped charges
is small. The calculated recombination rates, presented in Fig. \ref{fig:relaxation_rates},
demonstrate that while the effect of the dipole is weak, there is
still dependence on the separation of charges in the dipole: as the
separation is increased, the recombination rate is reduced. This is
due to the fact that the electrons and holes become condensed in separate
spatial regions so their overlap becomes smaller.

Similar result is obtained for the homogeneous recombination rate
$\gamma_{h}=$10 ns$^{-1}$ (100 ps decay time), which also
satisfies $\gamma_{h}<(\Delta k)^{2}D_{i}$. Since the rate
converges similarly and the distributions obtained are similar to
the previous case, we do not present charge distributions. Only the final
recombination rates are presented in Fig. \ref{fig:relaxation_rates}
by red crosses. We find that the relative dependence on dipole separation
is slightly stronger, but still the effect is  weak.

\begin{figure}
\begin{centering}
\includegraphics[width=1\columnwidth]{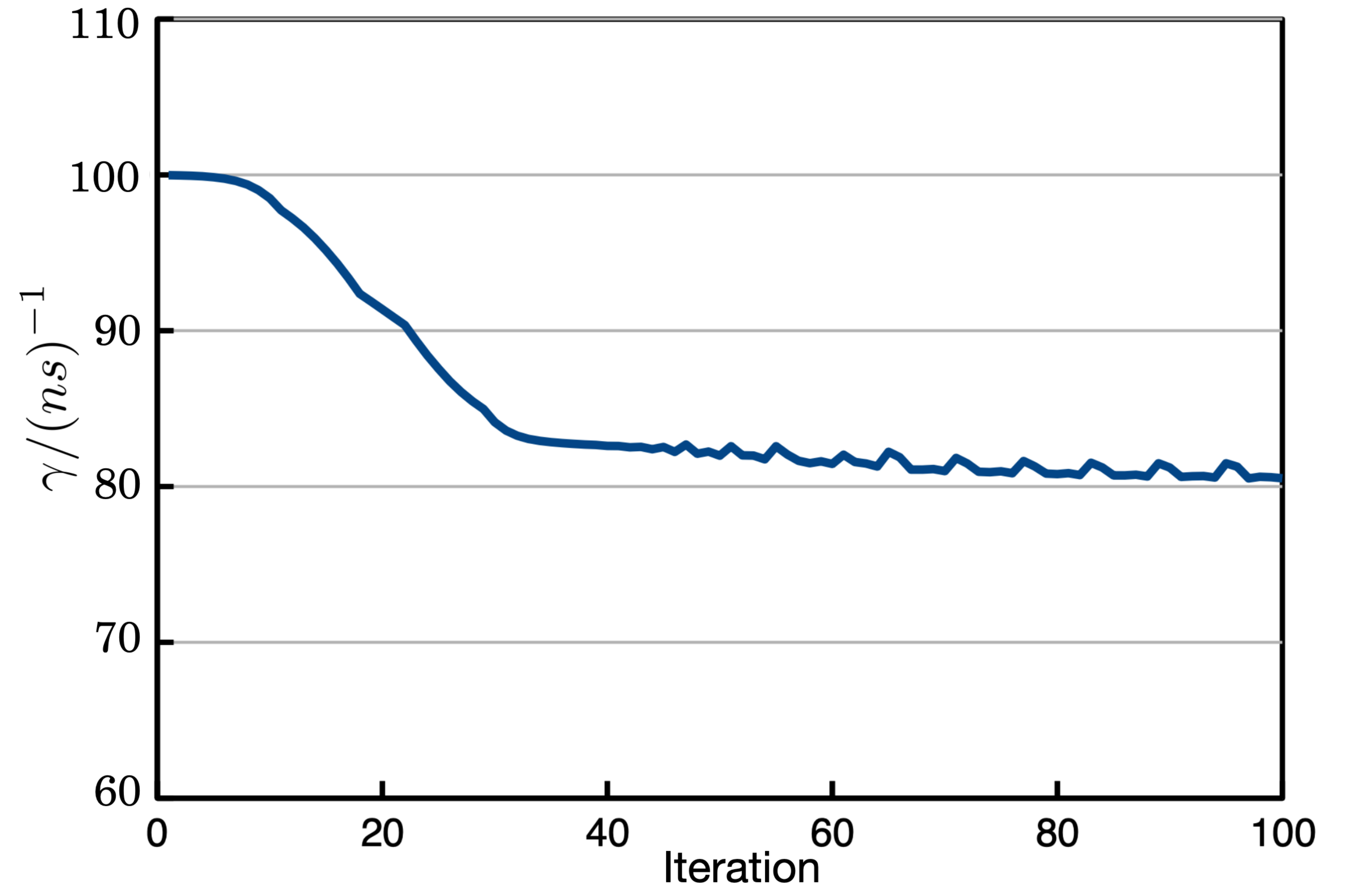}
\par\end{centering}
\caption{\label{fig:iteration-1}Convergence of charge distribution rate with
iterations. Dipole charges $Q=4e$, dipole separation $2z_{0}=4a\equiv100$
nm, homogeneous recombination rate $\gamma_{h}=100$ ns$^{-1}$.}
\end{figure}

\begin{figure}
\begin{centering}
\includegraphics[width=1\columnwidth]{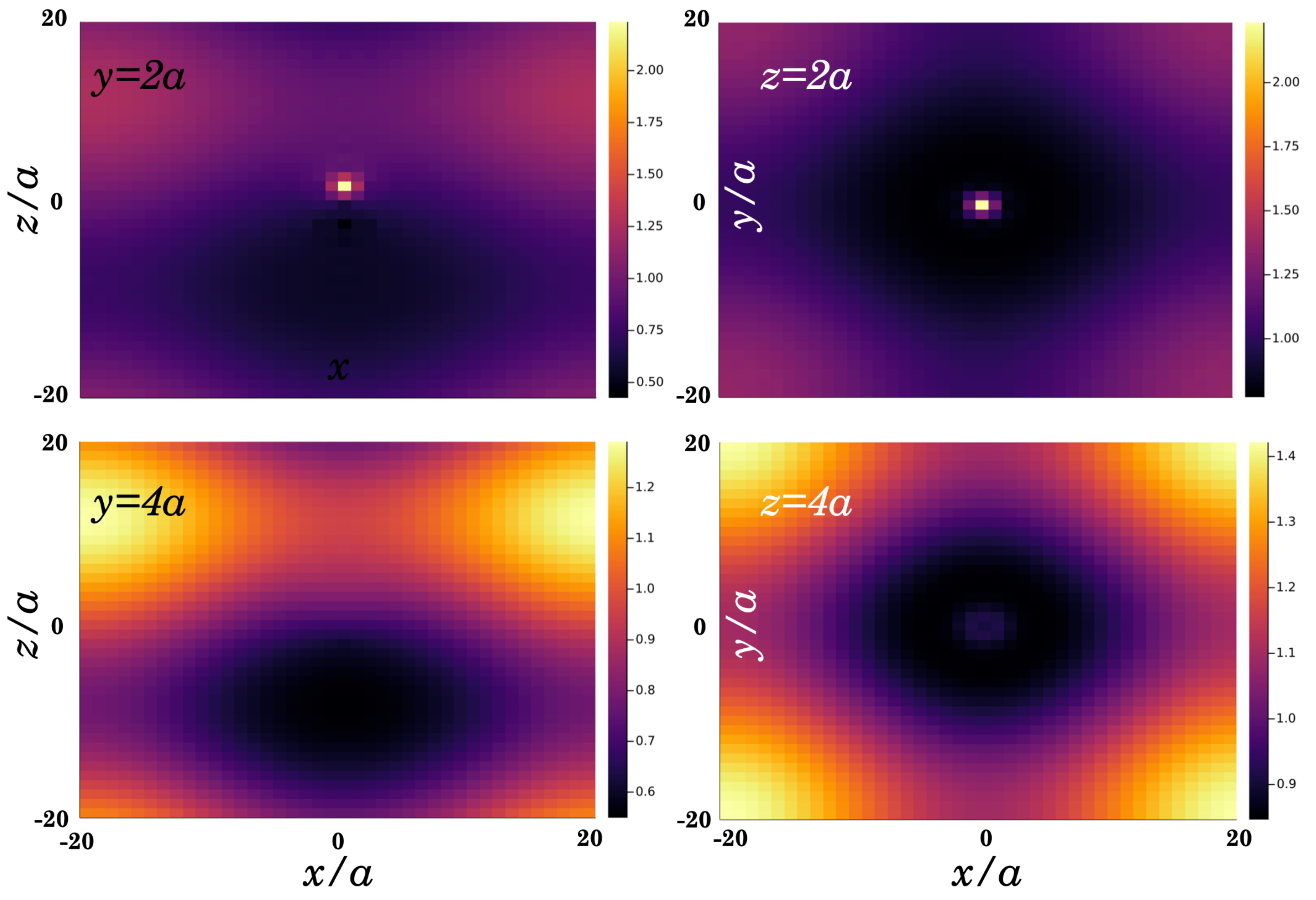}
\par\end{centering}
\caption{\label{fig:Distribution-of-electron-in-plain-1}Distribution of electron
density $\bar{\sigma}_{e}(\bm{r},t)$ at a specific plain. Dipole charges
$Q=4e$, dipole separation $2z_{0}=4a\equiv100$ nm, homogeneous recombination
rate $\gamma_{h}=100$ ns$^{-1}$. }
\end{figure}

\begin{figure}
\begin{centering}
\includegraphics[width=1\columnwidth]{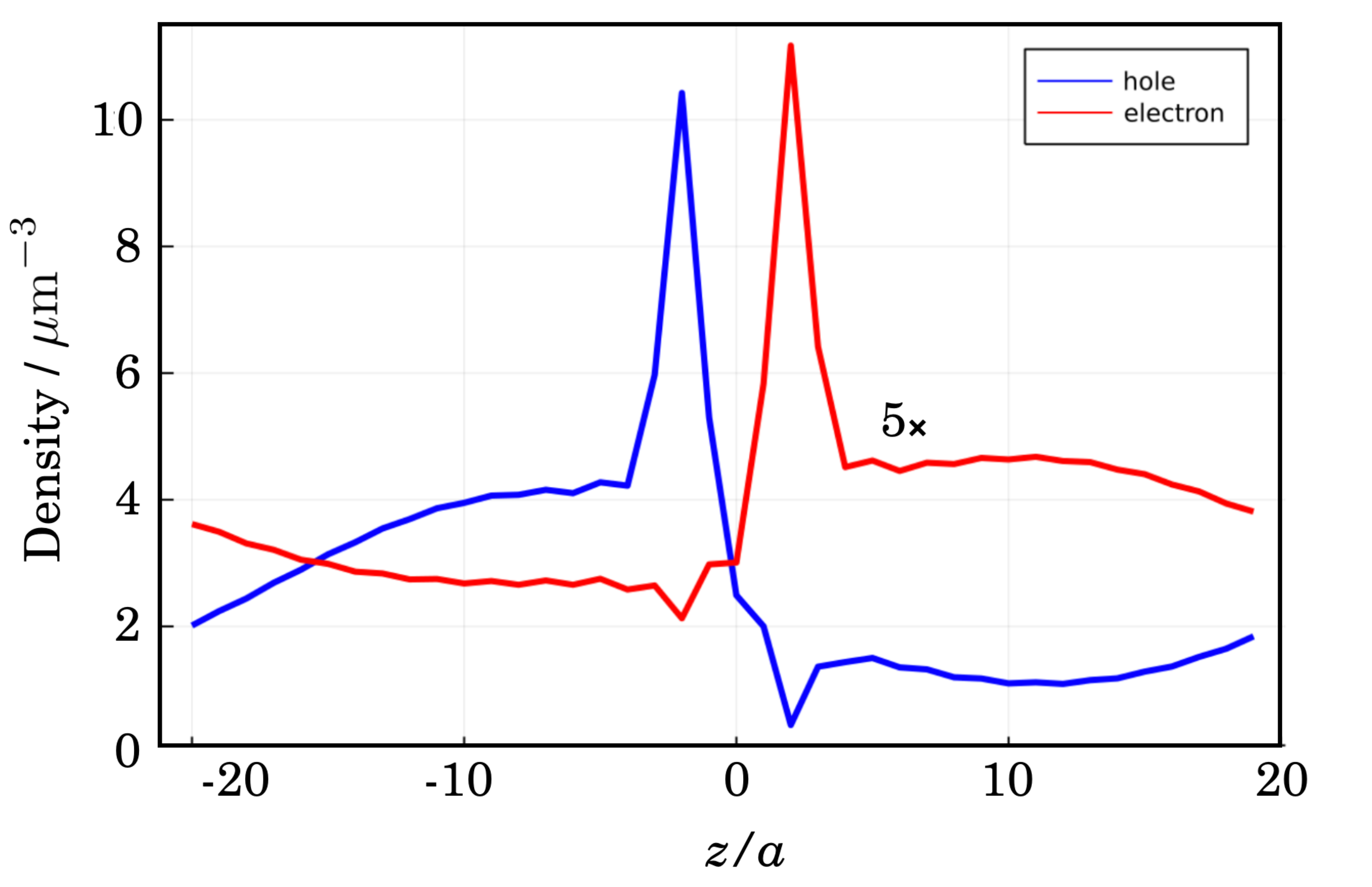}
\par\end{centering}
\caption{\label{fig:Distribution-of-electron-1}Distribution of electron (red)
and hole (blue) density $\bar{\sigma}_{e}(\bm{r},t)$ along $z$ axis at $x,y=0$.
Dipole charges $Q=4e$, dipole charge separation $2z_{0}=4a\equiv100$
nm, homogeneous recombination rate $\gamma_{h}=100$ ns$^{-1}$.}
\end{figure}

\begin{figure}
\begin{centering}
\includegraphics[width=1\columnwidth]{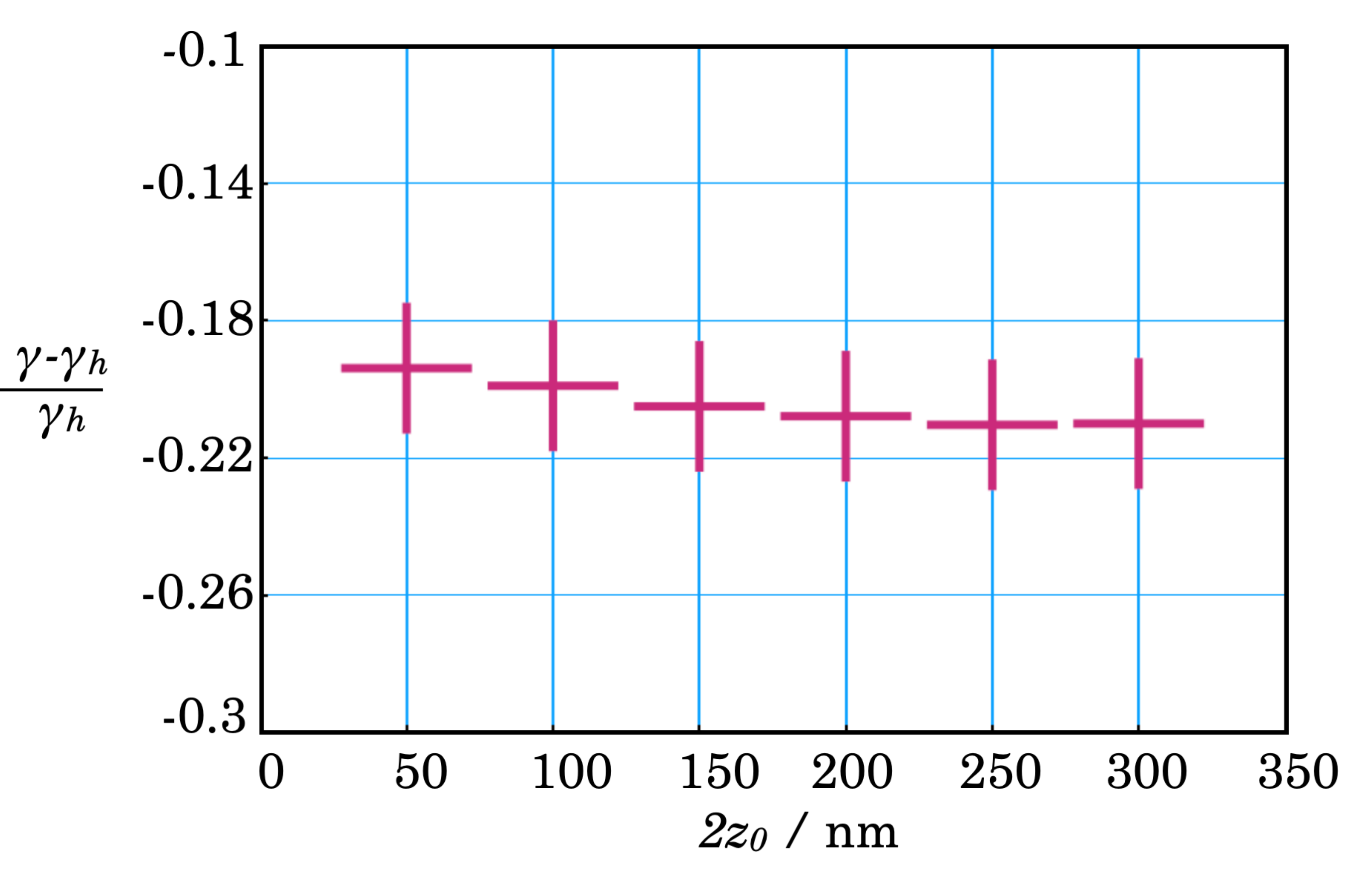}
\par\end{centering}
\caption{\label{fig:relaxation_rates-1}Calculated recombination rates for
dipole charges $Q=4e$ as a function of separation between dipole
charges when $\gamma_{h}=100$ ns$^{-1}$.}
\end{figure}

Completely different result is obtained when the homogeneous decay is fast:  $\gamma_{h}=100$ ns$^{-1}$ (10
ps decay time). In that case we have (for holes) the diffusion term
$\bm{k}^{2}D_{h}$ that can be larger or smaller than $\gamma_{h}$
depending on the amplitude of wave vector $\bm{k}$. In Fig \ref{fig:iteration-1}
we show the convergence of the solution. First of
all notice additional complexity in convergence. The dependence is
non-exponential - the rate evolves in step-like fashion as
$\gamma$ value traverses specific discrete values of $\bm{k}^{2}D$.
Due to numerical approach, the result does not relax to the complete
stable solution and small fluctuating behavior remains visible.
However, the rate can be estimated. For estimation of the rate we
take the lower limit, where the solution is mostly stabilized. 

Charge distribution becomes more localized compared to the previous case: 
in Fig. \ref{fig:Distribution-of-electron-in-plain} the distribution
was homogeneous except at the positions of trapped charges. In the
present case in Fig. \ref{fig:Distribution-of-electron-in-plain-1}
we find that the long-range charge attraction (due to positive charge)
and charge repulsion (due to negative charge) create depleted regions. That corresponds to
the situation where the charges recombine faster than they  escape
the dipolar wells. Very large variations of charge density are obtained
along $z$ axis as shown in Figs. \ref{fig:Distribution-of-electron-in-plain-1}
and \ref{fig:Distribution-of-electron-1}. The competition between
diffusion, recombination and drift can be observed. At $z>0$ the
positive defect charge attracts the electron density from the long
distances $z>0$ as the Coulomb potential is long-range. This shifts electron and hole distributions along $z$ axis in opposite directions so the recombination rate is reduced. Very close
to the defect positive charge (at $z\sim2a$), while charge density has spikes, 
 the opposite charges
are repelled out from the nearest neighborhood and the recombination
happens slowly.

Such large variation of electron density has a strong effect on the
recombination rate. In Fig. \ref{fig:relaxation_rates-1} we show
that the recombination rate is reduced from the homogeneous case by
20\% - much larger value compared to Fig. \ref{fig:relaxation_rates}.
The effect starts from smallest  $z_{0}$ values, while
for $2z_{0}>150$ nm dependence of the rate is almost constant. 

\section{Discussion and conclusions}

Presented model allows determining the 
effect of internal defect cluster fields on the recombination 
rate (or time). The model allows to assign a decay rate constant to the 
 highly non-stationary charge distribution decay process. 
 This becomes possible by using the finite box with periodic boundary conditions ($k$-space solution).  The approach, thus, allows to characterize the vicinity of the defect cluster. Considering the cluster without periodic boundary conditions it would be impossible to obtain a single rate constant by this method because the system would be described by non-exponential
 decay process  as the decay rate close to the cluster and far from it are essentially different. So the present model is approximate while allows to demonstrate the effect that local fields induce on the recombination process.

According to the presented calculations, the most significant parameter
describing charge evolution and recombination in the defect cluster
is the diffusion coefficient. It relates the diffusion length $l$
and diffusion time $\tau$ via relation $D_{i}=l_{i}^{2}/\tau_{i}$.
By taking the diffusion time to be equal to the lifetime, we obtain the total
diffusion length, i. e. 
the distance a charge can traverse during its lifetime by diffusion. 
The total
diffusion length is the parameter that should be correlated with the
size of the defect cluster. In our analysis we used a box with edge
length $L=$1 $\mu m$. If the total diffusion length 
is much longer than the defect cluster size, the charge will be able to escape
the cluster (and the Coulomb field of the dipole): the cluster
will have negligible effect on the charge recombination. Diffusion
constant for holes in silicon are of order of $10$ cm$^{2}$/s$\equiv 1000$ ($\mu$m)$^{2}$/$\mu$s,
consequently, the hole covers $\sim10$ ($\mu$m) before recombination
takes place if the  recombination time is $100$ ns. In this
case, if the defect cluster extends only over few hundred nm as it was assumed in simulations, its effect
on the recombination is almost vanishing, as is observed in Fig. \ref{fig:relaxation_rates}.
In the other limit, if the lifetime is short - few ps, the total diffusion length
corresponds to few tens of nm and that becomes comparable to the
size of the defect cluster, and the effect of the cluster dipolar field on
the recombination time  becomes considerable. 

 Defining the size of the defect cluster is also not trivial. As
Coulomb potential is long range, size of the cluster does not necessarily
corresponds with the distance between point defects in the cluster.
For the dipolar internal fields, the Coulomb potential energy of the
dipole decays approximately as $R^{2}$, where $R$ is the distance from the dipole.
Comparing the cluster-induced Coulomb potential energy (last two terms in Eq. \ref{eq:coulomb}) to $k_{B}T$ we can define
the effective dipolar capture distance, and the corresponding cluster
size $L_{d}=4z_0\sqrt{Q(8\pi\epsilon\epsilon_{0}z_0k_{B}T)^{-1}}$,
where $2z_0$ is the separation of dipolar charges. In our model for
dipole charge separation of $100$ nm and $Q=4e$ we can find $L_{d}\approx100$
nm. The relevant size of the cluster that should be considered for
the diffusion process is therefore even smaller than we have discussed
above. 

In this discussion we are not restrictive about  homogeneous system
recombination rate. The silicon crystals are never pure and various
types of impurities create recombination centers. The homogeneous
recombination rate therefore varies in various crystal preparation
conditions. Additionally, in order to isolate the effect of dipolar
fields we neglected the possibility that HE irradiation which creates
trapping centers (responsible for creation of internal fields), also
generates additional recombination centers in the same area. The local
 ``homogeneous'' recombination rate in the
vicinity of the defect clusters therefore could be quite different
compared to the bulk values of unaffected silicon. Therefore, high recombination rates in the vicinity of clusters and strong influence of 
the internal cluster fields on the recombination process is highly probable.

Diffusion process is quite easy to evaluate and estimate, while the
drift force depends on various additional parameters: the amount of
trapped charge, $Q$, effective separation of dipole charges (which
could be associated with the size of the defect cluster, as discussed
above). Consider holes as their diffusion constant is smaller.  The
quantity $\frac{4\pi\chi\mu_{i}}{V}\frac{Q}{e}$, which scales the
Coulomb fields in Eq. \ref{eq:real-space-final} (notice that $d^{3}k=(2\pi)^{3}/V$)
and  equal to $\frac{4\pi\chi D_{i}}{Vk_{B}T}\frac{Q}{e}=278\mu s^{-1}$
with $Q=4$ and at room temperature has the dimension on $s^{-1}$.
Parameter $\bm{k}^{2}D_{i}$ has the same dimension $s^{-1}$ and
 can be directly compared. Taking the smallest $\bm{k}\equiv dk=2\pi/L$
we find $\bm{k}^{2}D_{h}=47400\ \mu s^{-1}$. Diffusion coefficient
of electrons is even larger. The Coulomb drift forces acting on electrons
and holes are thus considerably weaker than the diffusion-related
gradient forces (including the large dielectric constant). This demonstrates
that the diffusion process is extremely fast in Si compared to microscopic
internal field-induced gradient forces due to trapped charges by defect
clusters. Additionally it should be taken into account that the Coulomb
field is long-range but quickly decays on small scale. The thermal
activation energy $k_{B}T$ overcomes the Coulomb attraction energy
for trapped charge $Q=4e$ at a distance of $20$ nm, hence, only
at such small distances the mobile charges cannot escape the potential
well of trapped charge. In our modeling the box size is $\sim1$ $\mu$m,
consequently, the large amount of charge is barely affected by Coulomb
attraction forces.

However, this study of a single dipolar internal field applies when
defect concentration is relatively small. If the defect concentration
is high, the internal fields are no more dipolar, bulk properties
like mobility, diffusion constant, effective masses may be extensively
affected. As the parameters become different, the eventual recombination
rate can be strongly affected. However, our model with the bulk diffusion constant being independent on the 
defect concentration allows to highlight how  the dipolar internal field
redistributes the mobile charges so that charge recombination becomes altered solely due to internal field effects.

Considering realistic samples we should keep in mind that high energy
irradiation creates clusters of defects with various configurations
of trapped charges. In Figs. \ref{fig:relaxation_rates} and \ref{fig:relaxation_rates-1}
we demonstrated how the rate of recombination changes with the simplest model of trapped  dipole. When the distance between trapped
charges increases, the whole picture would finally reflect
the case of independent positively and negatively charged point defects
scattered in the bulk. Since the governing parameter is the diffusion
length of a charge, the transition to independent charged defects
is directly related to the total diffusion length. 

The homogeneous recombination rate, that we
consider in our simulations is  barely observable quantity in experiments
due to variability of types of recombination centers in irradiated silicon.
As we have demonstrated the large effect of the dipolar field on recombination
rate emerges when the diffusion length is comparable with the dimension
of the defect cluster. This can be achieved when the size of the defect
cluster is large. As we have discussed, for the recombination rates
of $1$ $n$s, the effective diffusion length is $1$ $\mu$m. If
the defect clusters are of the same dimensions we will have conditions
that the charges cannot escape the internal defect fields during their
lifetime and the defect cluster is the determining factor governing
the recombination rate. Theoretically this is easily demonstrated by making order-by-order solution of the dynamic equation 
as demonstrated in Appendix \ref{sec:Order-by-order-solution}. 
When the dimensions of the cluster become comparable to the diffusion length (by varying parameter $\gamma_h$) 
the effect is amplified as difference $\bm{k}'^{2}D_i - \gamma_h$ can get to zero in the denominator.
Our results thus demonstrate that the ratio of the charge diffusion length and the size of clusters is the most important parameter that governs the sensitivity of the recombination rate to the defect density. 

Having this in mind, consider the case of independent homogeneously
distributed monopole trapped charges without clasterization, $i.$e. $z_{0}=\infty$ to the
case where defects are organized into clusters, i. e. $z_{0}$ finite.
Figures \ref{fig:relaxation_rates} and \ref{fig:relaxation_rates-1}
have to be compared starting from large $z_{0}$ and going to smaller
values. We then observe \emph{increasing} recombination rate with
dipolar-type  cluster formation. That is how the present model
should be conveyed. Totally pure semiconductor crystals do not exist
and recombination usually takes place via point defects even before
irradiation. In that case, the uniform distribution of point defects
would yield the homogeneous recombination rate of such sample $\gamma_{hom}$.
This parameter is essentially proportional to the density of the recombination
defects, $N_{r}$, while a single defect is characterized by a charge
capture radius, $R_{c}$. Irradiation of semiconductor by HE particles
creates clusters of defects with additional recombination centers
$M_{r}$ and trapping centers. Since no foreign atoms are introduced,
additional recombination defects are of the same type as the original
recombination defects, while their density is larger in the clusters.
We therefore should expect increase of $\gamma_{hom}$ in the vicinity
of the clusters, i. e. we can consider the  ``homogeneous''
recombination rate to take very different values in various spatial
positions. 

Another questionable point is whether bulk crystal parameters should
be applicable to the defect clusters. Charge mobility in Si decays
with the concentration of dopants (down to $\sim$100 cm$^{2}$/Vs
for electrons when defect concentration reaches $10^{19}$ cm$^{-3}$)\citep{Arora1982}.
 While overall concentration of defects may not be very high, local
concentration in the vicinity of defect cluster may be very high.
As we have discussed the lower diffusion coefficients means that variation
of the recombination rate at the cluster could be appreciable even
at low homogeneous relaxation rates $\gamma_{hom}$. Our simulations
thus correspond to the limiting case when density of defects inside
the cluster is low. We find that internal electric field of the cluster
becomes playing considerable effect at high concentration of point
recombination centers. If we assume that the charge mobility decreases
tenfold, the concentration of point recombination centers can be ten
times lower to observe the same strong effect. Electron mobilities
of $\sim$100 cm$^{2}$/Vs corresponds to the lower limit and thus
should represent the case when electrons diffuse over the point defects
via tunneling. So these conditions would correspond to the high concentration
of trapping centers and that would reflect the other limiting case.
However, in the conditions of high concentration of defects the structure
of internal fields should be also considered more complicated. 

It should be stressed that such clustering of defects may be 
important for various types of
semiconductor high energy irradiation sensors.
Defect clustering  creates local internal fields and conditions that are different from the bulk.
If they determine the bulk properties (e.g. by creating sinks of charge density)
this will be observed by bulk experiments.

Concluding, we have presented theoretical analysis of charge recombination
in Si crystal in the vicinity of trapped charge creating dipolar field
starting from the real space Fokker-Planck equation which describes
charge evolution including diffusion and drift forces as well as charge
recombination. Using the reciprocal space representation we have the
governing equation in the compact form,
what allows iterative solution of the equation. Using the stationary
conditions  we define the experimentally observed recombination
rate of the homogeneous system. We found that for slow recombination
process in the case of homogeneous conditions, inserting dipolar field
minimally affects the rate of homogeneous recombination. Large variation
of the recombination rate occurs when the homogeneous recombination
rate is fast (on the order of tens of picoseconds). The reason behind
this effect is that the diffusion length of charges during their lifetime becomes comparable
to the extensions of defect clusters.

\appendix

\section{\label{sec:Order-by-order-solution}Order-by-order solution}

In order to demonstrate the sensitivity of the effect on various parameters
it is instructive to perform analytically few steps. Hence, take Eq.
\ref{eq:final} and start from the homogeneous distribution 
\begin{equation}
\bar{\sigma}_{i}(\bm{k})=\left\{ \begin{array}{c}
1,\ \bm{k}=0\\
0,\ \mathrm{otherwise}
\end{array}\right.
\end{equation}
 The first iteration starting from homogeneous distribution we find
an additional term in distribution function at $\bm{k}\neq0$:
\begin{eqnarray}
\bar{\sigma}_{i}(\bm{k}) & = & \frac{(2\pi)^{3}}{V}\frac{iq_{i}Q\chi\tau_{i}}{\pi^{2}m_{i}^{\ast}\left(\bm{k}^{2}D_{i}-\gamma_{h}\right)}\sin\left(\bm{k}\bm{z}_{0}\right)\label{eq:whatever}
\end{eqnarray}
 which gives the recombination rate 
\begin{eqnarray}
\frac{\gamma-\gamma_{h}}{\gamma_{h}} & = & \frac{8Q^{2}\chi^{2}\tau_{h}\tau_{e}}{\pi Vm_{h}^{\ast}m_{e}^{\ast}}
\int
\frac{-d\bm{k}'\sin^{2}\left(\bm{k}'\bm{z}_{0}\right)}{\left(\bm{k}'^{2}D_{h}-\gamma_{h}\right)\left(\bm{k}'^{2}D_{e}-\gamma_{h}\right)}.\label{eq:rate_appendix}
\end{eqnarray}
 We thus find that at first iteration we obtain correction to the
recombination rate. Notice that $\gamma<\gamma_{h}$ if $D_{e}=D_{h}$.
However, the situation may become different if $D_{h}<D_{e}$ as in
the case of Si crystal. In that case the integral may yield negative
value if $\bm{k}'^{2}D_{h}<\gamma_{h}<\bm{k}'^{2}D_{e}$. In this
case the overall recombination rate may possibly become larger than
$\gamma_{h}$.

\section*{Acknowledgements} 
This work was performed in the framework of the CERN RD50 collaboration.
It was funded by the Lithuanian Academy of Sciences, grant No CERN-VU-2021-2022. Computations were
performed on resources at the High Performance Computing
Center, ‘‘HPC Sauletekis’’ in Vilnius University Faculty of Physics.

\section*{Author Declarations}
The authors have no conflicts to disclose.

\section*{Data Availability} 
The data that support the findings of this study are available from the corresponding author upon reasonable request.


%

\end{document}